\documentclass[preprint,showpacs,aps,a4paper]{revtex4}

\usepackage{graphicx}
\usepackage{dcolumn}
\usepackage{bm}
\usepackage[english,spanish,activeacute]{babel}
\def\journal#1#2#3#4{{\sl #1,\/} {\bf #2}, #3 (#4)}
\def\tur{\vphantom{|^{|^|}}}
\def\tdr{\vphantom{|_{|_|}}}
\begin{document}
\selectlanguage{english}

\title{Flux of atmospheric muons: Comparison between AIRES simulations
and CAPRICE98 data.}
\author{P.~Hansen, P.~Carlson, E. Mocchiutti}
\affiliation{Royal Institute of Technology (KTH),\\
             AlbaNova University Center \\
             S-10691 Stockholm,\\
             Sweden.}
\author{S.~J.~Sciutto}
\affiliation{Departamento de F\'isica\\
             Universidad Nacional de La Plata\\
             C. C. 67 - 1900 La Plata\\
             Argentina.}
\altaffiliation{Fellow of Consejo Nacional de Investigaciones
Cient{\'\i}ficas y T\'ecnicas of Argentina.}

\author{M. Boezio}
\affiliation{ University of Trieste and
Sezione INFN di Trieste \\
Via A.
Valerio 2\\
I--34147 Trieste\\
Italy}

\date{8$^{\rm th}$ July 2003}

\begin{abstract}

We report on a comparison between the flux of muons in the atmosphere
measured by the CAPRICE98 experiment and simulations performed with
the air shower simulation program AIRES.
To reduce systematic uncertainties we have used 
as input 
the primary fluxes of protons and helium nuclei also measured by 
the CAPRICE98 experiment.
 Heavy nuclei are also taken into
account in the primary flux, and their contribution to the muon
flux is discussed.  The results of the simulations show a very good
agreement with the experimental data, at all altitudes and for all
muon momenta. With the exception of a few isolated points, the
relative differences between measured data and simulations 
are smaller than 20 \%; and in all cases compatible with zero
within two
standard deviations. The influence of the input cosmic ray flux on the
results of the simulations is also discussed.  This report includes
also an extensive analysis of the characteristics of the simulated
fluxes.

\end{abstract}

\maketitle

\section{Introduction}

Detailed measurements and studies of the flux of muons in the
atmosphere represent a subject of particularly great interest. This is
mainly due to the fact that a measurement of the muon flux is an
indirect measure of the neutrino flux and can therefore be used to
improve the calculation of the atmospheric neutrino
flux which in turn is used to
compare with the observed neutrino rates in experiments targeted to
detect neutrino oscillations.  In addition, comparing measurements of
fluxes of muons and other particles at different altitudes with
simulated data is a powerful tool to check and/or calibrate air shower
simulation programs. Such programs are not only essential to predict
the atmospheric neutrino flux, but also play an important role in the
analysis of data taken at highest energy air shower experiments like
Auger \cite{AUGERW}, AGASA \cite{AGASA} or HiRes \cite{HiRes}.  They
are also used in \v Cerenkov light detector experiments like AMANDA
\cite{AMANDA} or ANTARES \cite{ANTARES} to estimate backgrounds
from atmospheric neutrinos.

Several calculations of the atmospheric muon flux at different altitudes
have been performed in the past \cite{naumov,fluka,ralph}. At ground
level more data is available allowing also for simultaneous absolute
flux and $\mu^+/\mu^-$ flux ratio analysis \cite{willi,wentzicrc,wentzprd}.
The comparisons with experimental data
show discrepancies in particular at low rigidities.
The most significant differences between simulated and
real data can be attributed to the following effects:

\begin{description}
\item[-] The normalization of the primary cosmic ray flux. 
Different measurements disagree in about 10 \% or more
on the proton and helium spectra between 10 and 50 GeV
\cite{naumov}.
\item[-] The solar modulation, the geomagnetic effect
and the model of the atmospheric profile. 
\item[-] The uncertainties in the particle 
production models, particularly the hadronic 
interaction generators 
~\cite{muonpaper}.
\end{description}

The recent introduction of Ring Imaging \v Cerenkov counters (RICH) in 
balloon-borne spectrometers has made it possible to measure
in the large proton background also positive muons.
The CAPRICE98 experiment with a gaseous RICH \cite{yo}, 
measured the flux of positive (negative) muons with
momenta in the range 0.3-20 GeV/$c$ (0.3-40 GeV/$c$), at
different altitudes ranging from ground level
(885 g/cm$^2$ atmospheric overburden) up to float altitude (5.5 g/cm$^2$).

In this work we make a comparison between the direct measurement of the flux
of muons in the atmosphere by the CAPRICE98 experiment~\cite{yo} and
the corresponding
simulated values obtained from the air shower simulation program AIRES
\cite{AIRESManual}.
This program is a 3D Monte Carlo simulator
where the majority of the  processes that may undergo the 
shower particles, 
are taken into account.  
AIRES also has the advantage of including effects of the curvature of the 
Earth and of the geomagnetic field.
The AIRES program has the possibility to swap 
between different hadronic models and it includes links to
two well-known external hadronic interaction packages, namely
SIBYLL \cite{SIBYLL} and QGSJET \cite{QGSJET}. 

This paper is organized as follows:
In section \ref{S:TheExperiment} we give some details of the
CAPRICE experiment.
In section \ref{S:TheSimulations} we report on the method used to
simulate the muon flux, using as input experimental values for the 
cosmic ray fluxes at the top of the atmosphere (TOA).
The results of the comparison between experiment and simulations are
reported in section \ref{S:Results} and finally we summarize our
conclusions in section \ref{S:Conclusions}.

\section{The experiment}
\label{S:TheExperiment}

The analysis presented in this work use the muon
fluxes measured by the CAPRICE98 
(Cosmic Anti-Particle Ring
Imaging \v Cerenkov Experiment) experiment, at all the atmospheric
depths given in table \ref{T:DepthIntervals} 
\cite{yo}.
A detailed description 
of the procedures for muon identification, a careful 
study of the efficiency of each detectors, the various
sources of background, and the rejection criteria and
surviving contamination, can also be found in \cite{yo}.

The balloon 
was launched from Ft. Summer,
New Mexico, USA ($34.28^{\circ}$N, $104.14^{\circ}$W) on the 28$^{th}$
May 1998 \cite{amb99}. 
The following measurements were performed: 
\begin{itemize}
\item \textbf{Muon measurement at ground}.
Before launch the spectrometer took 
data at ground level for a period of about  
14 hours. The ground level at Ft. Summer 
is located 
at an altitude of 1270 m that corresponds to an atmospheric 
depth of 885 g/cm$^{2}$.
\item \textbf{Muon measurement during balloon ascending period}.
During the ascent period of a few hours 
measurements were done on 
the flux of particles as a function of momentum and atmospheric depth.
\item \textbf{Muon measurement during balloon floating period}.
During the 
20 hours at float
above 35 km, corresponding to approximately 5
g/cm$^2$ of residual atmosphere, the balloon spectrometer 
recorded data on muons, protons and helium nuclei.
This last portion of the flight is important because
it provides the 
data used to estimate the primary flux of proton and helium nuclei 
at the top of the atmosphere 
used in the simulations.
\end{itemize}

At every altitude the experimental data available includes the
$\mu^{+}$ ($\mu^{-}$) flux for rigidities ranging from 
0.3 GV up to 20 (40) GV. It is important to stress that this
is the first experiment able to measure positive muons
up to 20 GV.
In the previous CAPRICE94 balloon flight \cite{boe99b,muonpaper}
the upper limit for $\mu^{+}$ was 2 GV.

The CAPRICE98 spectrometer accepts 
particles arriving with an inclination with 
respect to the vertical axis of less than 20$^{\circ}$. This 
characteristic of the instrument needs to be taken 
into account when performing a simulation.
It is important to point out that the axis of the spectrometer
remained vertical during the flight.

\begin{table}
\[
\begin{array}{ccc}
\hline
\hline\tur\tdr
\hbox{\bfseries\itshape Level} &
\multicolumn{2}{c}{\hbox{{\bfseries\itshape Depth range} (g/cm$^2$)}} \\
\hbox{\bfseries\itshape } &
\hbox{\bfseries\itshape min-max } &
\hbox{{\bfseries\itshape average\/}} \\
\hline\tur
ground & 885-885&885\\
10 & 581-885&704 \\
9 &  380-581&462 \\
8 &  250-380& 308\\
7 &  190-250&219 \\
6 &  150-190&165 \\
5 &  120-150&136 \\
4 &   90-120&104 \\
3 &   65-90&77 \\
2 &   33-65&48.4 \\
1 &   15-33&22.6 \\
float & 5.45-5.95&5.5\tdr \\
\hline
\hline
\end{array}
\]
\caption{Atmospheric depth intervals defined for CAPRICE 98.}
\label{T:DepthIntervals}
\end{table}

\section{The simulations}
\label{S:TheSimulations}

\subsection{Flux at the top of the atmosphere}
\label{S:fluxtoa}

The main input for the simulation of the flux of atmospheric muons at
a given altitude is the absolute flux of cosmic rays at the top of the
atmosphere. In the calculation we included
 the
fluxes of the following 11 cosmic nuclei: H (protons and
deuterium), He (He$^3$ and He$^4$), C, N, O, Ne, Mg, Si, and Fe.

The most important contribution to the total absolute flux at the top
of the atmosphere comes from hydrogen and helium nuclei with only
small contributions from other 
nuclei. 
It is also important to mention that photons and electrons do not
contribute significantly to the flux of muons 
and therefore have not been included in our input.

We have mainly used  hydrogen and helium fluxes obtained by the CAPRICE98
experiment \cite{emiliano},
 thus ensuring that the bulk of the input
flux is affected by similar systematic errors as all the secondary
particles, in particular muons, since all of them are measured with
the same apparatus. This implies that when comparing experimental data
at a given atmospheric depth with the corresponding simulations, a
direct evaluation of the properties of the propagating algorithms is
being performed, minimizing the uncertainty due to any
inaccuracies in the input flux.

The CAPRICE98 experiment measured the absolute flux of
cosmic protons (helium nuclei) with kinetic energies ranging from
3 to 350 GeV (0.9 to 170 GeV/nucleon)
 \cite{emiliano}. These absolute fluxes
can be adequately parameterized as polynomial functions of the
logarithm of the primary energy. The deuterium to proton and He$^3$ to
He$^4$ ratios were taken from reference
\cite{isotopo}.   
For heavier nuclei we have used data 
from reference \cite{gordo}. 

\begin{figure}[p]
\begin{center}
\includegraphics{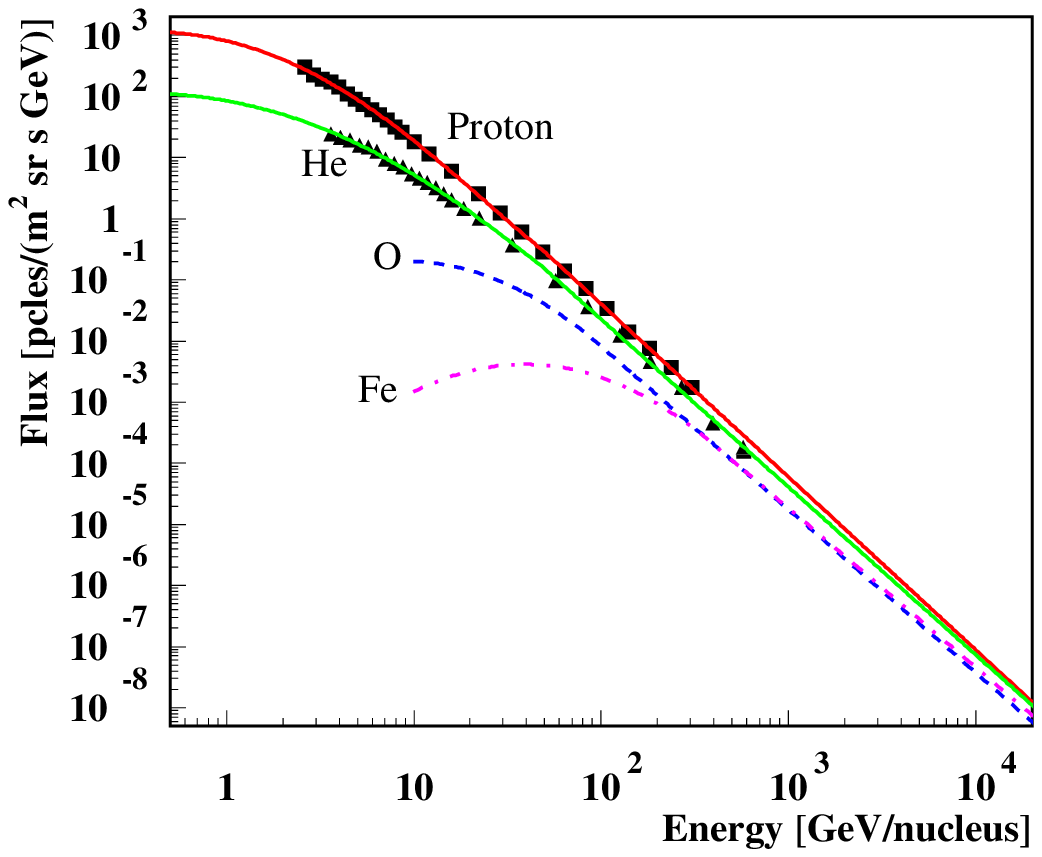}
\caption{Absolute fluxes of proton, helium, carbon, oxygen, and iron nuclei at
the top of the atmosphere, plotted as functions of the kinetic energy of the
primary. The points represent data from the CAPRICE98
experiment \cite{emiliano}, in the cases of proton (squares) and
helium nuclei (triangles).
The lines are the polynomial functions discussed in the text.}
\label{fig:fluxestoa}
\end{center}
\end{figure}

 The absolute TOA fluxes of hydrogen
(proton + deuterium), helium (He$^{3}$ +
He$^{4}$), carbon, oxygen, and iron are shown as functions
of the energy of the primary particles in figure~\ref{fig:fluxestoa}.
 The lines represent the
polynomial functions used as input for the simulations. The CAPRICE98
proton and helium data are also shown.
The parameterized
representations agree excellently with the experimental data.

It is important to remark that for low energies (below about 3 GeV) the solar
modulation was taken into account when fitting the plotted functions,
considering the level of solar activity registered at the date of the
CAPRICE98 flight.

Notice also that the input fluxes used in the simulations correspond
to total absolute fluxes at TOA without any correction due to
geomagnetic effects. Such effects are taken into account by the
simulation and analysis chain used to evaluate the different simulated
observables: a corrective weight is applied to the secondaries
generated by each primary entering the atmosphere to effectively take
into account the geomagnetic cutoff. Then, the particles are
propagated within the atmosphere, also taking into account their
deflections due to the geomagnetic field, which is in this case taken
as constant.

\subsection{Air shower simulations}

 The AIRES simulation engine \cite{AIRES,AIRESManual} provides full
space-time particle propagation in a realistic environment, taking
into account the characteristics of the atmospheric density profile
(using the standard US atmosphere \cite{atmosfera}), the Earth's
curvature, and the geomagnetic field (calculated for the location and
date of the CAPRICE98 flight with an uncertainty of a few percent
\cite{GF}).

The following particles are taken into account in the AIRES simulations:
photons, electrons, positrons, muons, pions, kaons, eta mesons, lambda
baryons, nucleons, antinucleons, and nuclei up to $Z=36$. 
Nucleus-nucleus, hadron-nucleus and
photon-nucleus inelastic collisions
with significant cross-sections are taken into account in the
simulation. The hadronic processes are
simulated using different models, accordingly to the energy: high
energy collisions are processed invoking an external package (SIBYLL
2.1 \cite{SIBYLL} or QGSJET01 \cite{QGSJET}), while low energy ones are
processed using an extension of Hillas splitting algorithm (EHSA)
\cite{AIRES,AIRESCCICRC,AUGERJPG}. The threshold energies separating
the low and high energy regimes used in our simulations are 200 GeV
and 80 GeV for the SIBYLL and QGSJET cases, respectively. The EHSA low
energy hadronic model used in AIRES is a very fast procedure,
effectively emulating the major characteristics of low energy hadronic
collisions. The model is adjusted to retrieve similar results as the
high energy hadronic model for energies near the transition thresholds
previously mentioned, and the low energy cross sections are calculated
from parameterizations of experimental data. A complete discussion on
the low energy hadronic models is clearly beyond the scope of this
paper. A separate report on this subject will be published elsewhere
\cite{sjsinpreparation}.

 AIRES has been successfully used to study several characteristics of
high energy showers, including comparisons between hadronic models
\cite{AUGERJPG,doqui99}, influence of the LPM effect \cite{LPM}, muon
bremsstrahlung \cite{mubrem}, and geomagnetic deflections \cite{GF} on
the shower development. AIRES has also been used to obtain an energy
calibration of the AGASA experiment \cite{AGASAc}, and to study the
expected efficiency of the Auger Observatory for detecting
quasi-horizontal showers generated by $\tau$-neutrinos
\cite{Pierretau}.

In the present work, AIRES has been used to simulate showers
with primary energies from 7.5 $\times$ 10$^{8}$
up to 10$^{15}$ eV.
In order to accurately simulate the absolute fluxes already described in
section \ref{S:fluxtoa}, and also to optimize the statistics, 
we conveniently divided the primary energy range into many
subintervals with boundaries chosen so as to have at each
of them approximately constant compositions and slope $\gamma$
($\gamma=d\Phi/dE_{pr}$ where $\Phi$ is the flux and $E_{pr}$
is the primary energy).
The independent sets of simulated showers were generated for 
each one of the intervals, considering also an isotropic 
arrival direction distribution with zenith angles ranging
from 0$^{\circ}$ to 89$^{\circ}$.

The shower simulations performed for this study add up to more than
300 millions of showers, generating particle data files with a total
size of about 30 GB, and requiring about 20 days of processing time
(using a 1 GHz processor).
The generation of such large set of simulated showers
proved to be a straightforward computing exercise, where the 
AIRES system could be easily configured for this particular
task, even if it was originally designed to simulate showers with
significantly larger energies.

When configuring the simulation program, several aspects have been
taken into account to properly set up the input parameters in our case
of flux simulation. In particular, we would like to make the following
remarks:
(i) The low energy hadronic interactions increase their importance
as the primary energy decreases. For this reason, we have done a
careful setting of the parameters of the EHSA, taking into account
experimental results, and comparisons with other models.
(ii) The statistical sampling algorithm of AIRES (the so-called
thinning) \cite{AIRESManual} was completely disabled. This means that
all the secondaries
generated during the shower development are fully propagated.
(iii) All electromagnetic particles (photons, electrons, and positrons)
with energies below 100 MeV were discarded. This significantly reduces
the processing time required for a given simulation, without altering
the propagation of hadrons and muons.

\subsection{Simulating the flux of secondary particles.}

Let us consider a given observing level located at an altitude
$h$ smaller than the injection altitude $h_{i}$.
After simulating the shower in the conditions 
of section  \ref{S:fluxtoa}, the secondary particles arriving to the
observing level, are processed
to estimate the corresponding fluxes, according to the following 
conditions:

\begin{description}
\item[-] When comparing with CAPRICE98 measurements \cite{yo}, 
only particles reaching the observing level with an inclination 
of less than 20$^{\circ}$ have been considered. In this paper all
particles satisfying that selection criterion
are referred as {\em quasi-vertical\/} particles.
 Unless otherwise specified all the results discussed in the following
 sections apply to quasi-vertical particles. Notice also that {\em
   full\/} fluxes refer to fluxes of particles coming from all directions.
\item[-] When necessary, the selected secondary particles are binned
according to their momenta, using the momentum intervals of the
experimental data.
\item[-] The selected particles are weighted to take into account the absolute
flux normalization associated with the primary energy subinterval 
(see section \ref{S:fluxtoa}) that corresponds to the shower
being analyzed. 
\end{description}
 
The procedure to evaluate the fluxes is repeated at each 
of the observing altitudes that are considered, using at each altitude
an independent set of simulated showers.

\section{Results}
\label{S:Results}
\subsection{General analysis of the simulated flux}

\begin{figure}[p]
\begin{center}
\includegraphics{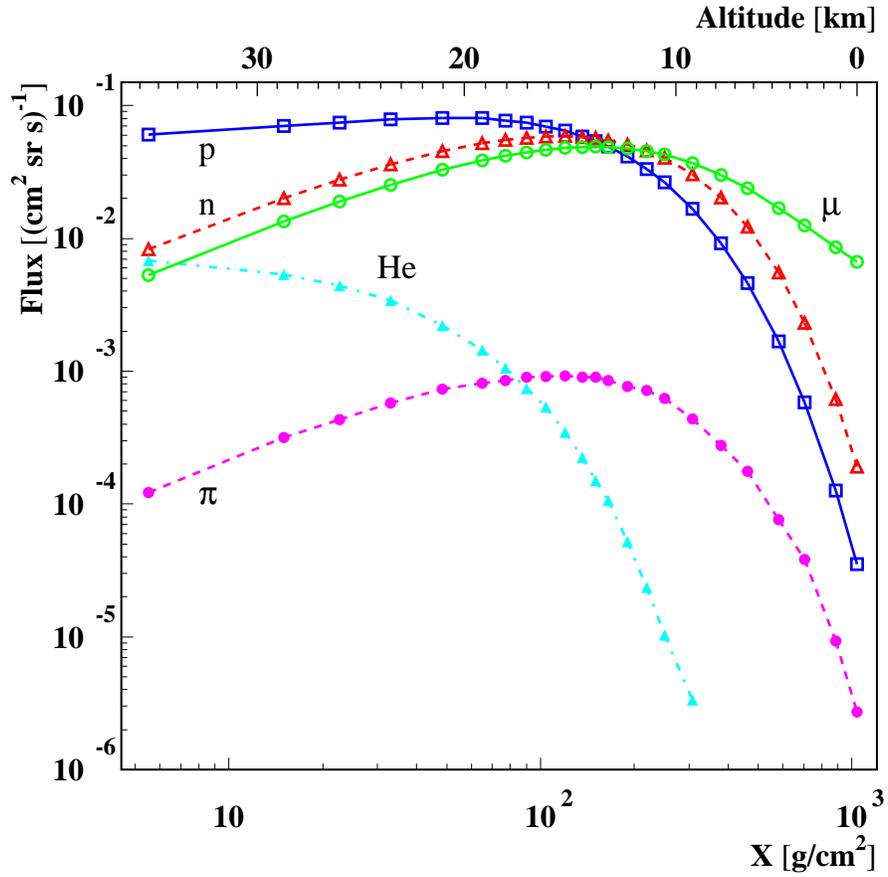}
\caption{Total fluxes of muons, pions, protons,
neutrons and helium nuclei as a function of atmospheric
depth and altitude above sea level.
The lines are drawn to guide the eye. 
} 
\label{fig:allabsfluxesvsxh}
\end{center}
\end{figure}

The fluxes of secondary particles have been simulated for all 
altitudes listed in table \ref{T:DepthIntervals}.  We have also
performed simulations at sea level (1035 g/cm$^{2}$).  In figure
\ref{fig:allabsfluxesvsxh} the total fluxes of muons, pions, protons,
neutrons and helium nuclei are plotted as function of atmospheric
depth and altitude above sea level.  Electrons, positrons and photons
were not propagated in detail in the simulation (see section
\ref{S:TheSimulations}) and are therefore not considered in our study.
Some of the main characteristics of these particle fluxes show up in
this figure.  At the highest level (5.5 g/cm$^2$) protons are the most
abundant particles because only 
a small fraction of them have interacted. 
  On the other hand, at
altitudes near sea level, the total flux is dominated by muons, which
account for more than 96 \% of the considered particles.  Notice also
that, as expected, the muon flux increases with decreasing altitude,
until reaching a maximum around
$X=150$ g/cm$^2$. 
The muon flux then decreases with decreasing altitude until
sea level.  For atmospheric depths in the range 90 -- 300
g/cm$^2$ the flux of muons is quite constant.  The pion
flux behaves similarly, reaching its maximum approximately at the same
altitude as that for the muons.

\begin{figure}[p]
\begin{center}
\includegraphics{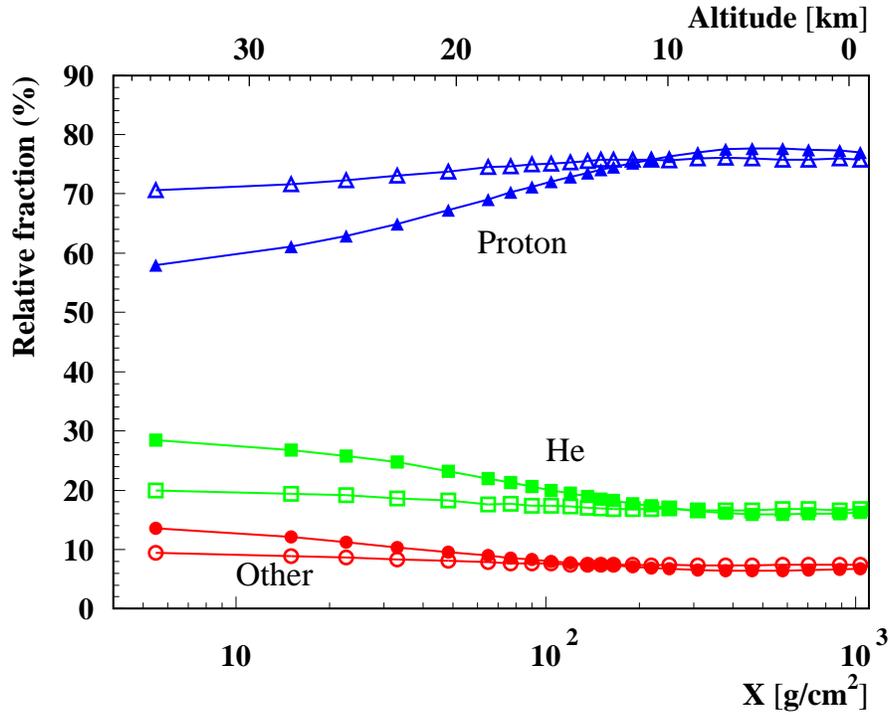}
\caption{Relative contributions of different 
primary particles to the full muon flux as function
of atmospheric depth and altitude above sea level.
The solid (open) symbols correspond to all muons (muons with momentum
greater than 10 GeV/$c$).
The lines are drawn to guide the eye.
} 
\label{fig:pcontribvsx}
\end{center}
\end{figure}

We have studied how the different primaries that make up 
the primary flux at the TOA, already described in section
\ref{S:fluxtoa}, contribute to the total muon flux at different
altitudes.
In figure \ref{fig:pcontribvsx} the relative contributions
of proton, helium nuclei, and other heavy nuclei
(C, N, O, Ne, Mg, Si, Fe) to the muon flux are plotted 
versus atmospheric depth
and altitude above sea level. It is evident that protons
give the largest contribution at all altitudes, and their
contribution increases when the altitude decreases.
The heavy nuclei contribute with a small but not completely
negligible fraction that ranges from about 8 \% at sea
level up to more than 10 \% at $X=5.5$ g/cm$^{2}$.

The contributions to the flux of muons with momenta larger than 10
GeV/$c$ is also displayed (open symbols) in figure
\ref{fig:pcontribvsx}. The different fractions are similar to the
corresponding ones for the total muon flux, except at high altitudes
where the proton fraction is substantially larger in comparison with
the previous case.

\begin{figure}[p]
\begin{center}
\includegraphics{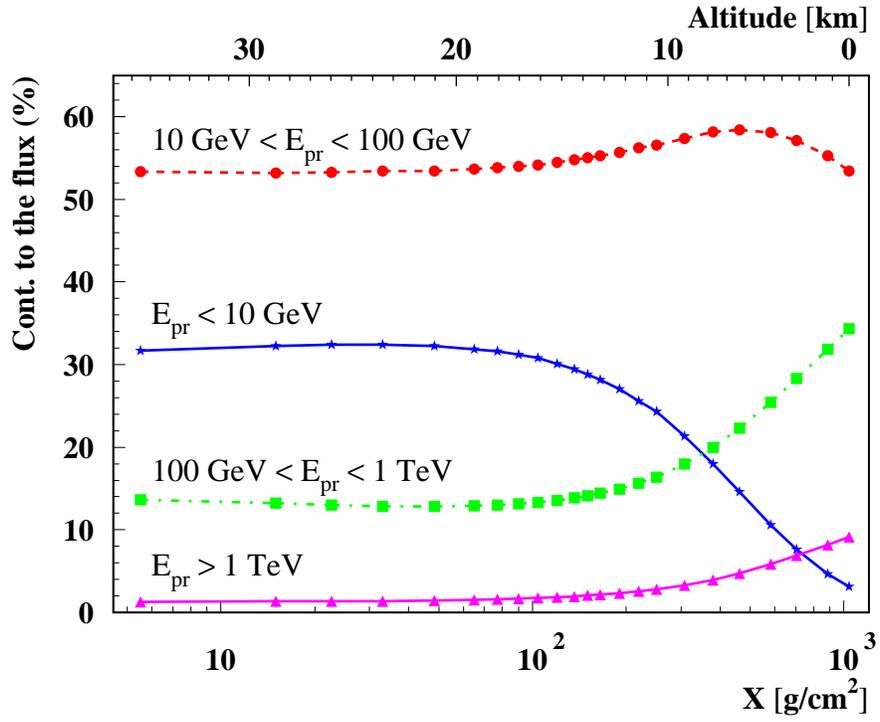}
\caption{Relative contribution 
of primaries with different energies to the full muon flux as
function of atmospheric depth and altitude above sea level.
The lines are drawn to guide the eye.
} 
\label{fig:contribpegyfluxmuvsxh}
\end{center}
\end{figure}

It is also interesting to analyze how primaries with different energies
(for nuclei total kinetic energy)
contribute to the muon flux.  To this end, we have divided the
primary energies $E_{pr}$ into four ranges, namely, (1) $E_{pr}$ $<$ 10
GeV; (2) 10 GeV $<$ $E_{pr}$ $<$ 100 GeV; (3) 100 GeV $< E_{pr} < 1$
TeV; and (4) $E_{pr}$ $>$ 1 TeV. The
corresponding contributions to the muon flux using the data of our simulations 
are displayed in figure \ref{fig:contribpegyfluxmuvsxh}, where
the relative contributions of each of the four primary energy
categories to the full muon flux 
(including all arrival directions)
are plotted versus atmospheric depth and altitude.  As expected,
in the energy range 10-100 GeV primaries, that account for the most significant fraction
of particles capable of entering into the Earth's atmosphere, are the
ones that contribute most to the muon flux at all altitudes. Lower
energy primaries contribute significantly (30 to 40 \%)
only at high altitudes, but their contribution decreases rapidly for $X$
longer than
100 g/cm$^2$. Near sea level, the
contributions of energy range 3 and 4 primaries increases significantly 
with a contribution from 100 GeV -1 TeV primaries of more than
30  \% (20 \% when selecting only quasi-vertical muons).  We have studied the
contribution of very low energy primaries ($E_{pr}$ $<$ 3 GeV) (the
results have not been included in figure
\ref{fig:contribpegyfluxmuvsxh} for simplicity).  Their contribution
to the flux is always small ($<1.5$\%) in the entire range of
altitudes considered.  This implies that any errors in the
estimation of the input flux for very low energies are unlikely to
have a significant impact on the simulated muon fluxes (notice that 3
GeV is the lowest primary energy measured by CAPRICE98).

\begin{figure}[p]
\begin{center}
\includegraphics{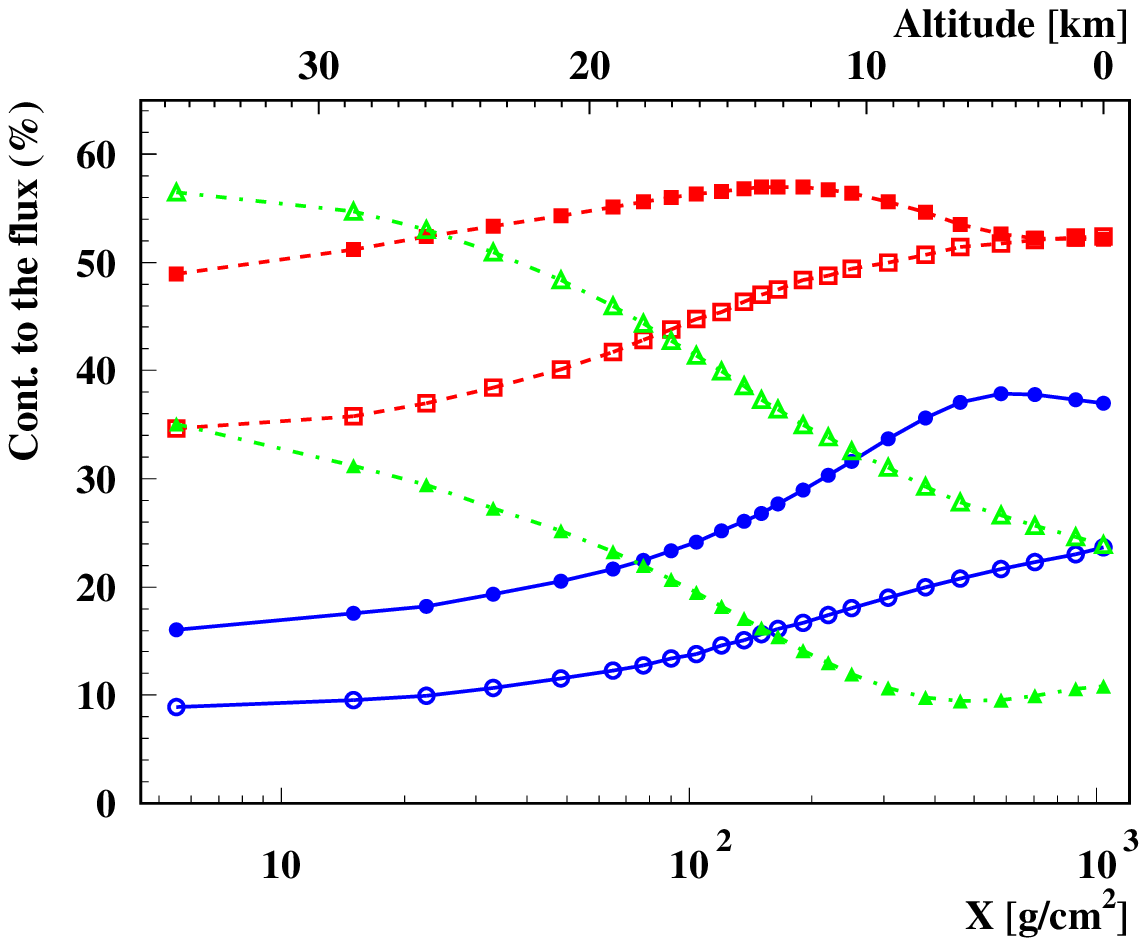}
\caption{Relative contribution 
of primaries with different zenith angles to the full muon flux as
function of atmospheric depth and altitude above sea level.
For zenith angle:
$\Theta$ $<$ 30$^\circ$ (circles),
30$^\circ$ $<$ $\Theta$ $<$ 60$^\circ$ (square),
60$^\circ$ $<$ $\Theta$ (triangle).
The solid (open) symbols correspond to all muons (muons with momentum
greater than 10 GeV/$c$).
The lines are drawn to guide the eye.
} 
\label{fig:contribzenfluxmuvsxhcone80deg}
\end{center}
\end{figure}

\begin{figure}[p]
\begin{center}
\includegraphics{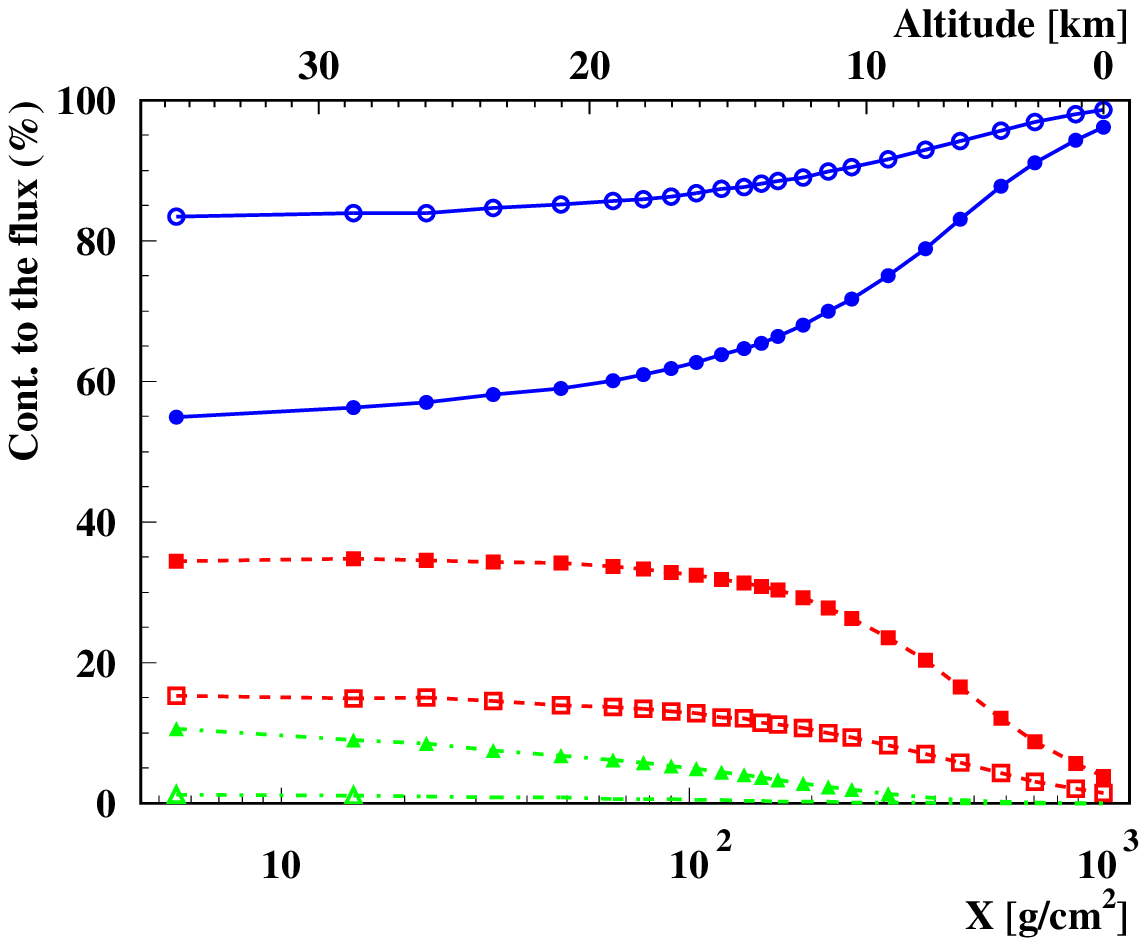}
\caption{Relative contribution 
of primaries with different zenith angles to the flux of
quasi-vertical muons as
function of atmospheric depth and altitude above sea level.
For zenith angle:
$\Theta$ $<$ 30$^\circ$ (circles),
30$^\circ$ $<$ $\Theta$ $<$ 60$^\circ$ (square),
60$^\circ$ $<$ $\Theta$ (triangle).
The solid (open) symbols correspond to all muons (muons with momentum
greater than 1 GeV/$c$).
The lines are drawn to guide the eye.
} 
\label{fig:contribzenfluxmuvsxhcone20deg}
\end{center}
\end{figure}

To complete our study of how the different primary particles
contribute to the flux of muons at different altitudes, we have
analyzed the number of muons generated from primaries with given
zenith angles. The results of this analysis are displayed in figures
\ref{fig:contribzenfluxmuvsxhcone80deg} and
\ref{fig:contribzenfluxmuvsxhcone20deg}.

In figure \ref{fig:contribzenfluxmuvsxhcone80deg} the relative
contributions to the full flux of muons corresponding to primaries
with zenith angle, $\Theta$, less than 30$^\circ$ (circles), greater than
30$^\circ$ and less than 60$^\circ$ (squares), and greater than
60$^\circ$ (triangles), are plotted versus atmospheric depth and
altitude. The solid (open) symbols correspond to all muons (muons with
momentum larger than 10 GeV/$c$).

The curves in this figure present a complicated dependency on 
altitude of the corresponding contributions. At sea level, showers
with $30^\circ < \Theta < 60^\circ$ dominate the  contribution 
of the full flux, and the different  contributions are
independent of the muon momentum. At high altitudes the contribution of
very inclined showers is dominating. This is due to the fact
that such showers pass through a thicker layer of air and are
more developed and producing more muons before reaching
observation level, in comparison with vertical ones
(see the discussion on angular distributions of muons
later in this section). This effect is more important in the case of
high energy muons, as indicated by the open symbol curves.

This picture changes dramatically when considering only quasi-vertical
muons, as displayed in figure
\ref{fig:contribzenfluxmuvsxhcone20deg}. In this case the showers with
$\Theta < 30^\circ$ are the ones that most contribute at all
altitudes, and their relative contribution is larger for high energy
particles, reaching virtually 100 \% when selecting muons with momenta
greater than 10 GeV/$c$. Notice that in figure
\ref{fig:contribzenfluxmuvsxhcone20deg} the open circles correspond to
muons with momenta larger than 1 GeV/$c$.

Another very important observable that we have considered in our
 analysis, is the distribution of arrival directions of muons at
 different altitudes.  As we have already commented in section
 \ref{S:TheExperiment}, and as explained in detail in reference
 \cite{yo}, the normalization of the muon fluxes measured by CAPRICE98
 is calculated under the assumption of an isotropic distribution of
 muons within the acceptance cone (zenith angle less than 20$^{\circ}$).
 In order to check the validity of this assumption, we have recorded
 angular distributions of muons at all the simulated levels and, 
 simultaneously, calculated the fluxes using several aperture cones.

\begin{figure}[p]
\begin{center}
\includegraphics{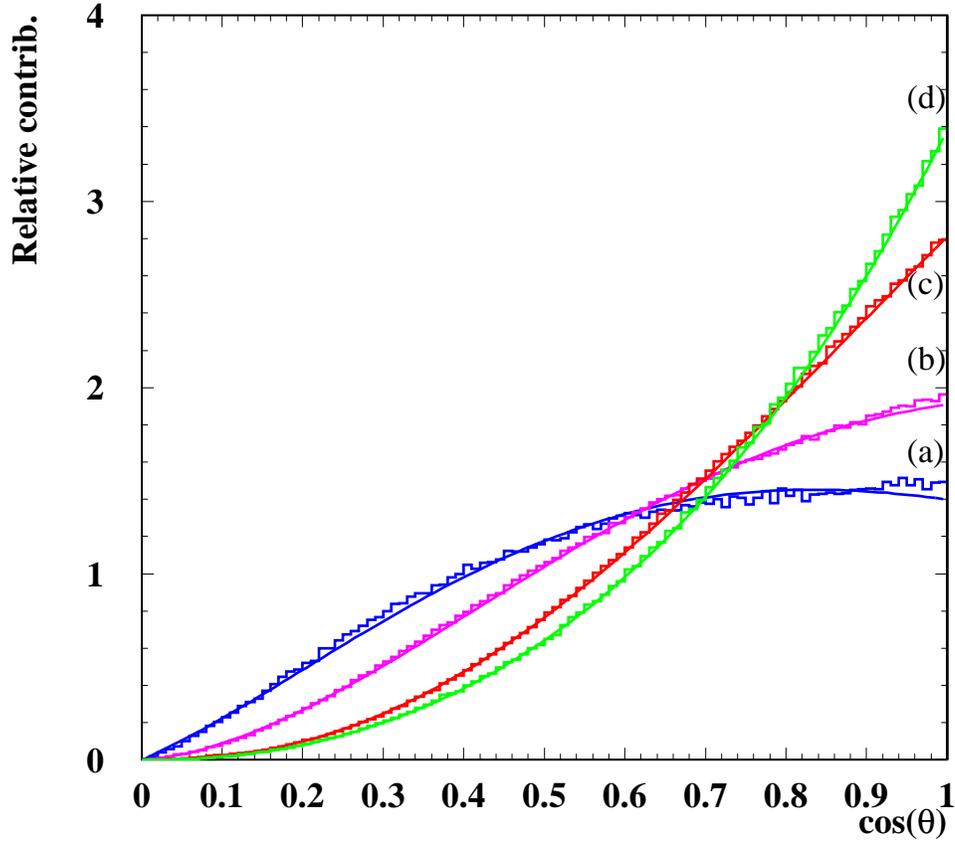}
\caption{Normalized angular distributions of muons
versus the cosine of the arrival zenith angle 
for several atmospheric depths:
(a) 5.5 g/cm$^{2}$, (b) 77 g/cm$^{2}$,
(c) 308 g/cm$^{2}$, and (d) 885 g/cm$^{2}$.
The histograms represent the simulated data while the smooth lines
correspond to fits of distribution to the form (\ref{equ:fit}).
20$^{\circ}$ correspond to $\cos\theta$=0.94.
} 
\label{fig:angledistribgfa}
\end{center}
\end{figure}

To start with, let us consider the angular distributions
of all atmospheric muons. In figure \ref{fig:angledistribgfa} 
the normalized angular distributions of muons are 
plotted versus the cosine of the arrival zenith angle in the entire
$\lbrack 0,1 \rbrack$ range for several altitudes namely (a) 5.5 g/cm$^{2}$,
(b) 77 g/cm$^{2}$,(c) 308 g/cm$^{2}$, and (d) 885 g/cm$^{2}$.

We have found that in the entire range of altitudes considered the
angular distributions can be accurately fitted by the  
following function:
\begin{equation}
f(\cos\theta)=U\, (\cos\theta)^{(\alpha + \beta \cos\theta)}
\label{equ:fit}
\end{equation}
where $\alpha$ and $\beta$ are constants, and $U$ is a normalization
factor determined by the condition $\int_{0}^{1} f(x) dx=1$. This
function can be conveniently fitted, with $\alpha$ and $\beta$ as free
parameters, using simulated or experimental data.

\begin{figure}[p]
\begin{center}
\includegraphics{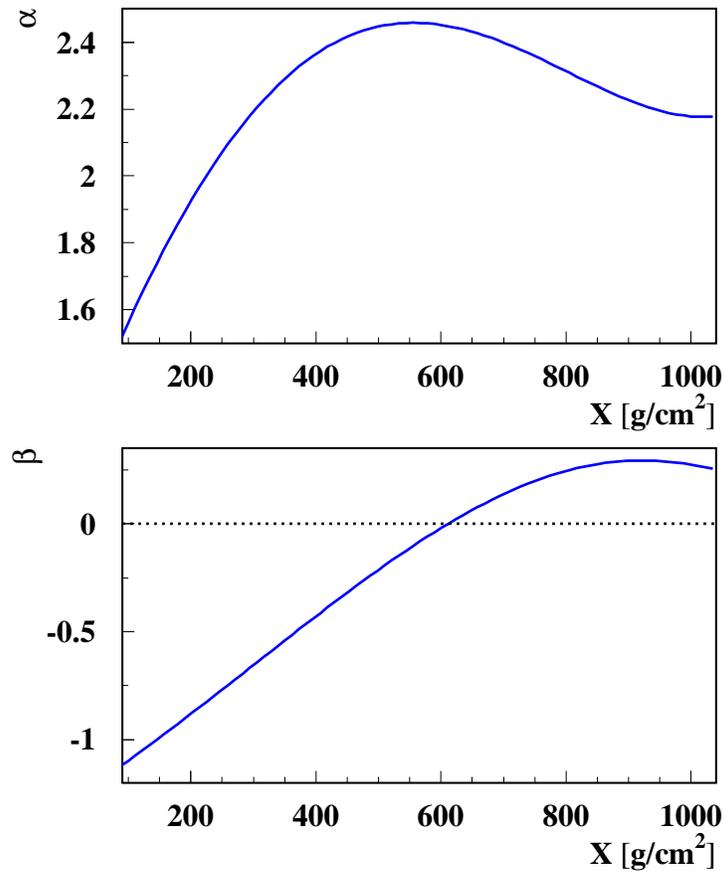}
\caption{$\alpha$ and  $\beta$
versus atmospheric depth.
} 
\label{fig:angledistparsvsx}
\end{center}
\end{figure}

The constants $\alpha$ and $\beta$ vary slowly as 
function of the atmospheric depth of the observing level
as illustrated
in figure \ref{fig:angledistparsvsx}.
Notice that for $x\sim600$ g/cm$^{2}$ $\beta \cong 0$,
and therefore in this case the distribution reduces to a power of
$\cos\theta$.

An important characteristic of the distributions plotted in figure
\ref{fig:angledistribgfa} is that
in the range $\theta < 20^{\circ}$ (corresponding to  $\cos\theta > 0.94$)
the variations as functions of angle are small, in particular
for high altitudes.
 This implies that the isotropy 
hypothesis is reasonably
justified and at the same time, the fluxes estimated using aperture
cones smaller than, say, 20$^{\circ}$, should not differ significantly.

\begin{figure}[p]
\begin{center}
\includegraphics{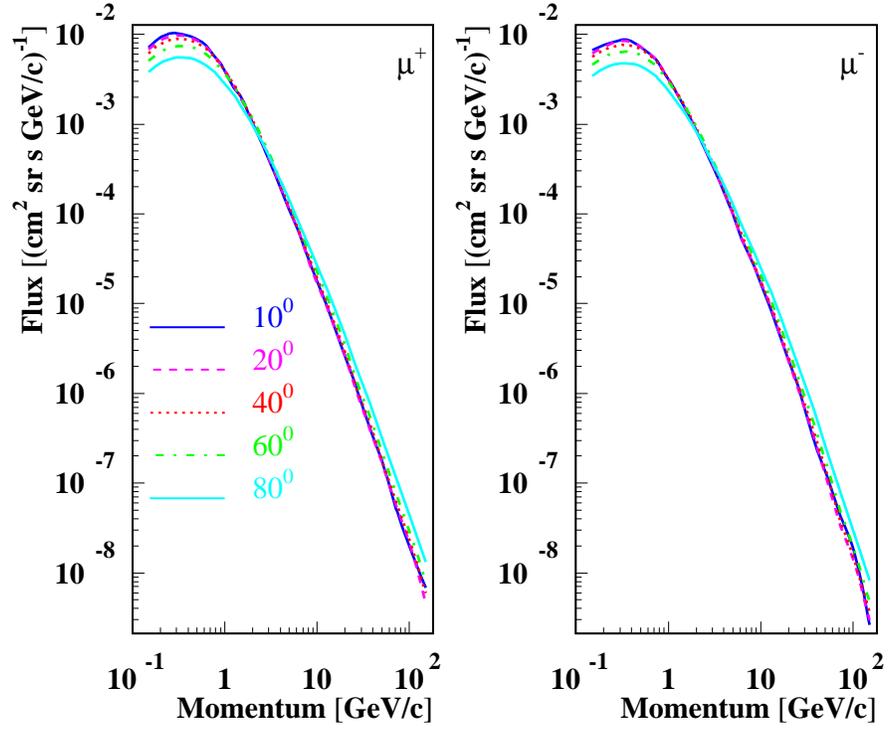}
\caption{
Simulated $\mu^{+}$ and $\mu^{-}$
fluxes at 
22.6 g/cm$^{2}$ calculated using different 
acceptance cones, plotted versus muon momentums.
} 
\label{fig:binfluxvscone22.6}
\end{center}
\end{figure}

\begin{figure}[p]
\begin{center}
\includegraphics{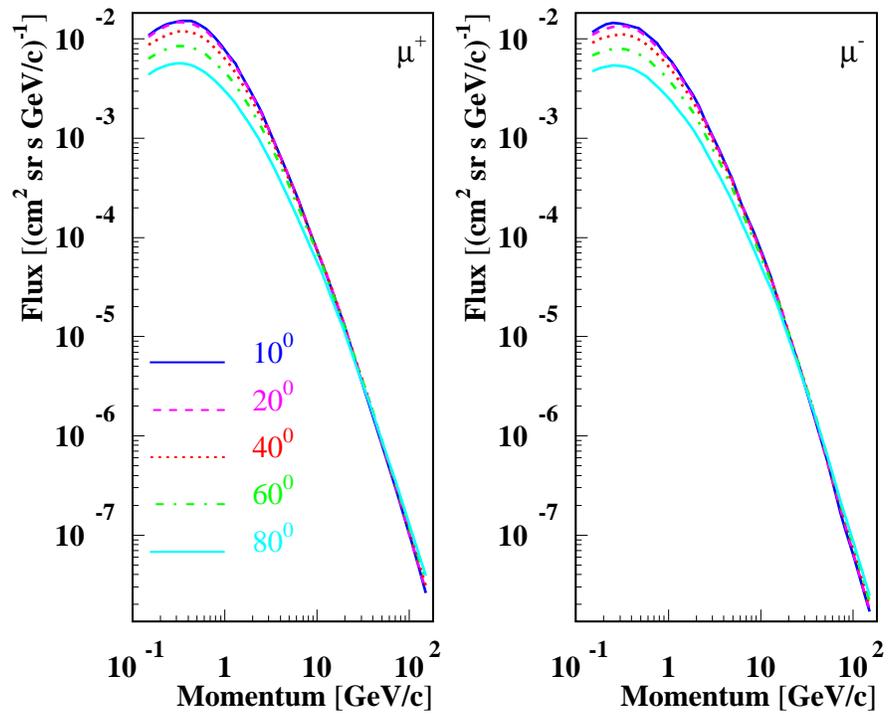}
\caption{ 
Same as figure \ref{fig:binfluxvscone22.6}, but for 
$x=308$ g/cm$^{2}$
} 
\label{fig:binfluxvscone308}
\end{center}
\end{figure}

\begin{figure}[p]
\begin{center}
\includegraphics{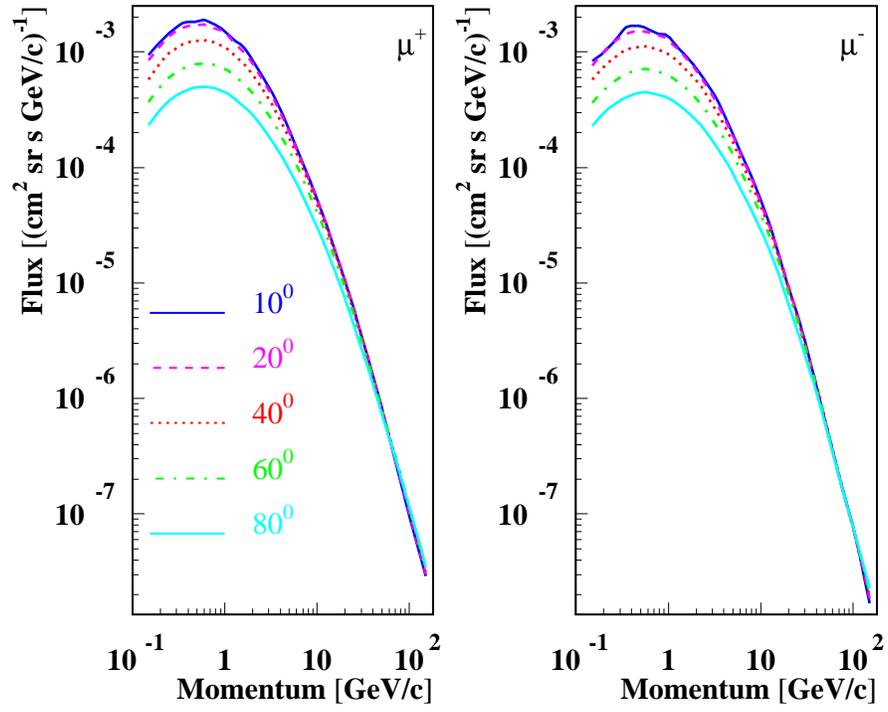}
\caption{ 
Same as figure \ref{fig:binfluxvscone22.6}, but for 
$x=885$ g/cm$^{2}$
} 
\label{fig:binfluxvscone885}
\end{center}
\end{figure}

This property shows up clearly from figures
\ref{fig:binfluxvscone22.6}, \ref{fig:binfluxvscone308}
and \ref{fig:binfluxvscone885}
 where the simulated $\mu^{+}$ and $\mu^{-}$
fluxes calculated using different acceptance cones, are plotted versus
muon momentum at atmospheric depths 22.6, 308, and 885 g/cm$^{2}$,
respectively.
Notice that there are no important differences between the 10$^{\circ}$ 
and 20$^{\circ}$ cases.

\begin{figure}[p]
{\centering \begin{tabular}{cc}
\resizebox*{0.46\textwidth}{!}{\includegraphics{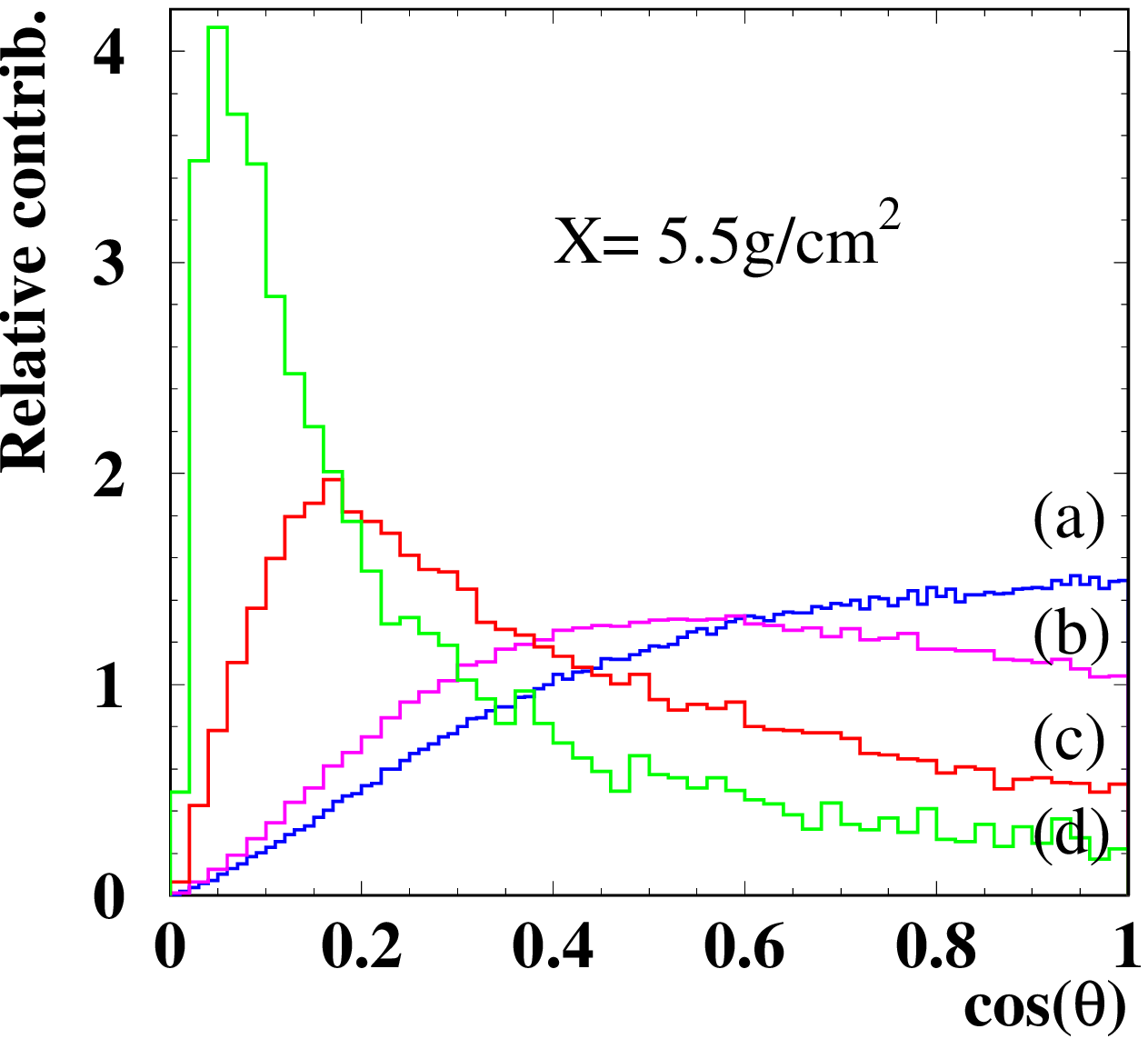}}
&\resizebox*{0.46\textwidth}{!}{\includegraphics{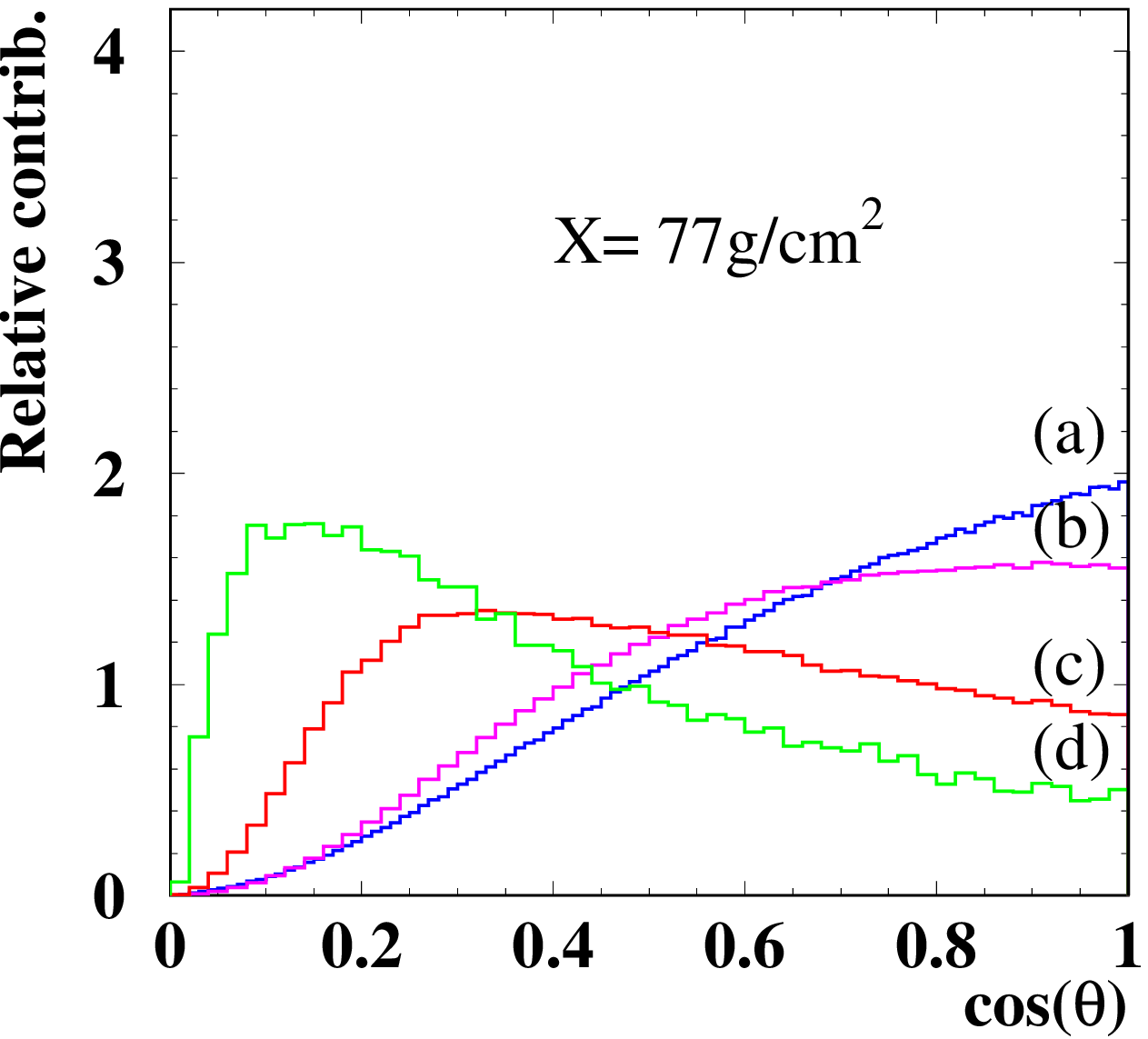}}
\\
\resizebox*{0.46\textwidth}{!}{\includegraphics{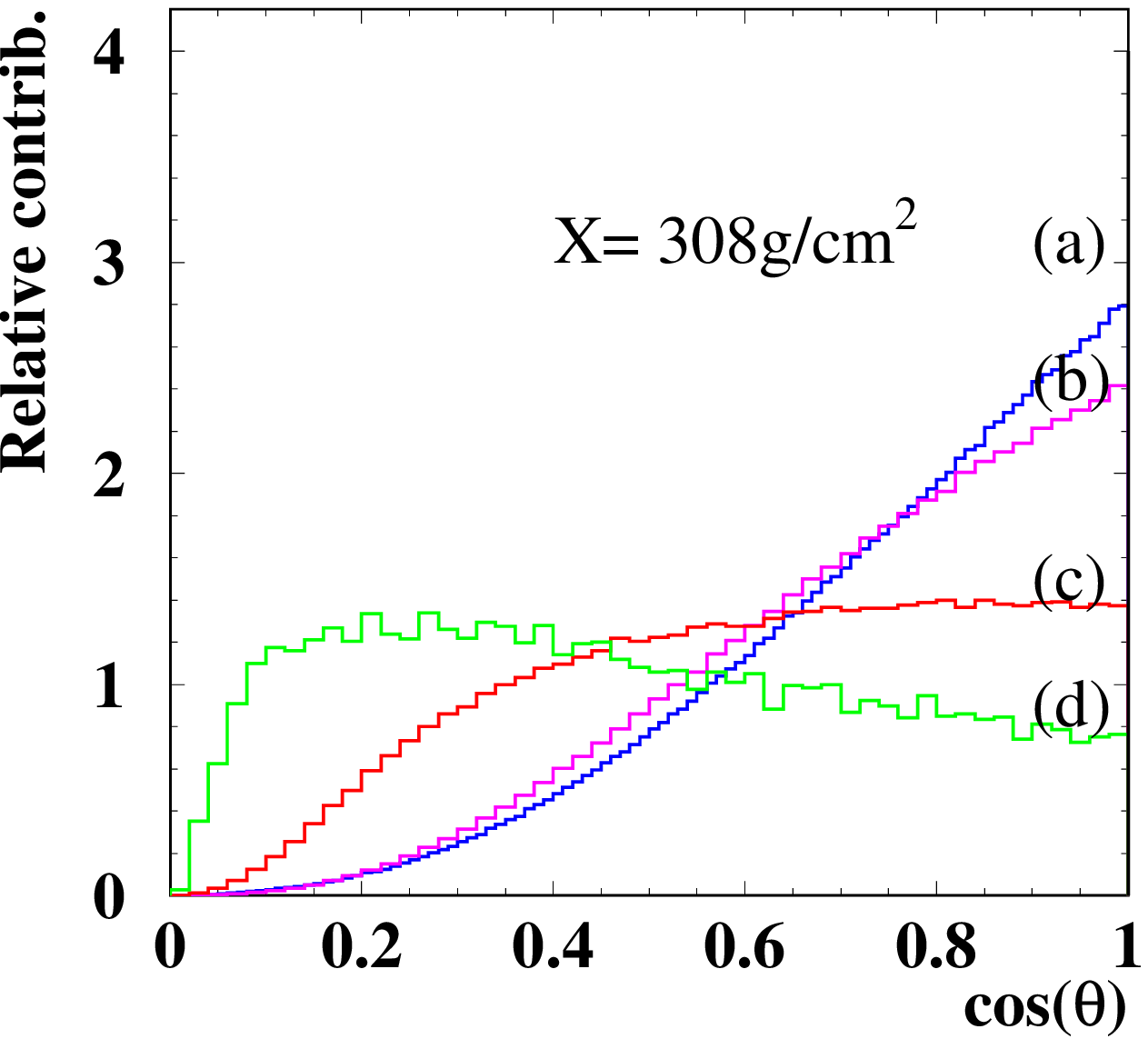}}
&\resizebox*{0.46\textwidth}{!}{\includegraphics{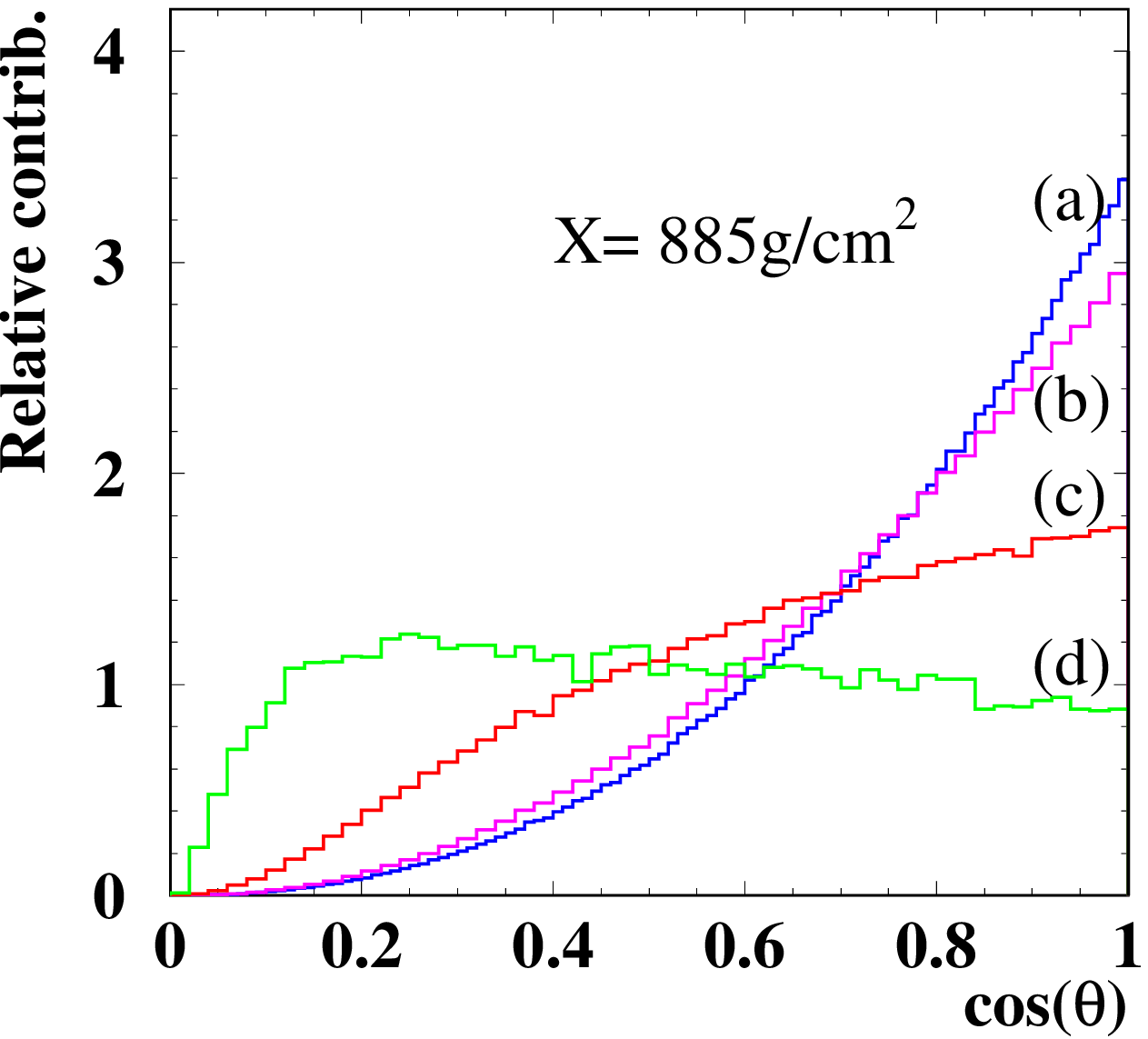}}
\\
\end{tabular}\par}
\caption{Normalized angular distributions of muons
versus the cosine of the arrival zenith angle
for several atmospheric depths.
The different  lines correspond to  
(a) all simulated muons, 
(b) $p_{\mu} > 1$ GeV/$c$,
(c) $p_{\mu} > 10$ GeV/$c$, and
(d) $p_{\mu} > 100$ GeV/$c$.
} 
\label{fig:ufa}
\end{figure}

From the distributions in figures \ref{fig:binfluxvscone308}
and \ref{fig:binfluxvscone885} it is clear that 
the high momentum end of the spectra do not 
significantly depend on the aperture cone used. 
However, this is not the case for the distributions in figure 
\ref{fig:binfluxvscone22.6} where it can be seen that the
steepness of the distributions for large momenta decreases with aperture
angle. This can be explained analyzing the angular distributions of
high energy atmospheric muons. In figure \ref{fig:ufa}, 
the angular distributions for
(a) all simulated muons, (b) $p_{\mu} > 
1$ GeV/$c$, (c) $p_{\mu} > 10$ GeV/$c$, (d)
$p_{\mu} > 100$ GeV/$c$,
are shown for four different 
altitudes.
We conclude from these figures that (i) the distributions
for the lower altitudes, 5.5 and 77 g/cm$^{2}$ 
show a large variation in angular distribution for different energies,
(ii) at high altitudes $\theta=0$ does no more
correspond to the maximum of the distribution (d)
which is located rather close to  $\theta=90^{\circ}$.
Therefore, when enlarging the aperture angle an increasing 
number of muons is accepted, consequently producing a 
larger flux at high momenta.

It is worthwhile mentioning that this feature of the angular
distribution of high energy muons at large altitudes implies that
any experiment having a narrow acceptance cone centered at 
the vertical only accumulates a very small fraction
of the muon flux.

\subsection{Comparison with experimental data}
\label{S:compex}

\begin{figure}[p]
\begin{center}
\includegraphics[width=0.8\textwidth]{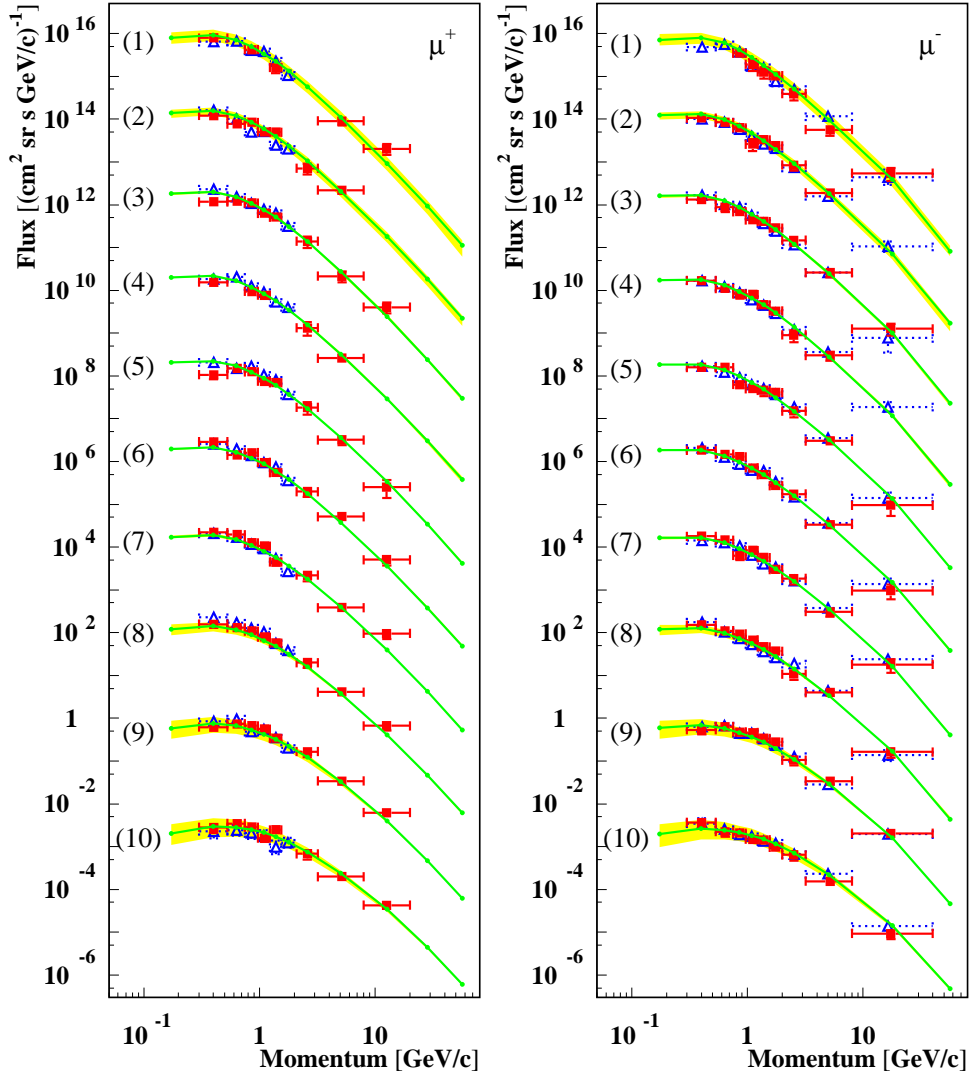}
\end{center}
\caption{
Absolute muon flux versus muon momentum. The solid squares 
(open triangles) correspond to CAPRICE98 (CAPRICE94)
data. The solid lines represent the AIRES simulations
at the respective average depths. The shaded areas
indicate the variations registered in the simulated flux
when passing from the minimum to the maximum depth of the 
corresponding
measurement interval.
Different curves represent different altitudes:
{\em (1)\/} 22.6 g/cm$^2$, scaled by $10^{18}$.
{\em (2)\/} 48.4 g/cm$^2$, scaled by $10^{16}$.
{\em (3)\/} 77 g/cm$^2$, scaled by $10^{14}$.
{\em (4)\/} 104 g/cm$^2$, scaled by $10^{12}$.
{\em (5)\/} 136 g/cm$^2$, scaled by $10^{10}$.
{\em (6)\/} 165 g/cm$^2$, scaled by $10^{8}$.
{\em (7)\/} 219 g/cm$^2$, scaled by $10^6$.
{\em (8)\/} 308 g/cm$^2$, scaled by $10^4$.
{\em (9)\/} 462 g/cm$^2$, scaled by $10^2$.
{\em (10)\/} 704 g/cm$^2$, scaled by 1.
}
\label{fig:binfluxvsp9498ascen}
\end{figure}

\begin{figure}[p]
\begin{center}
\includegraphics{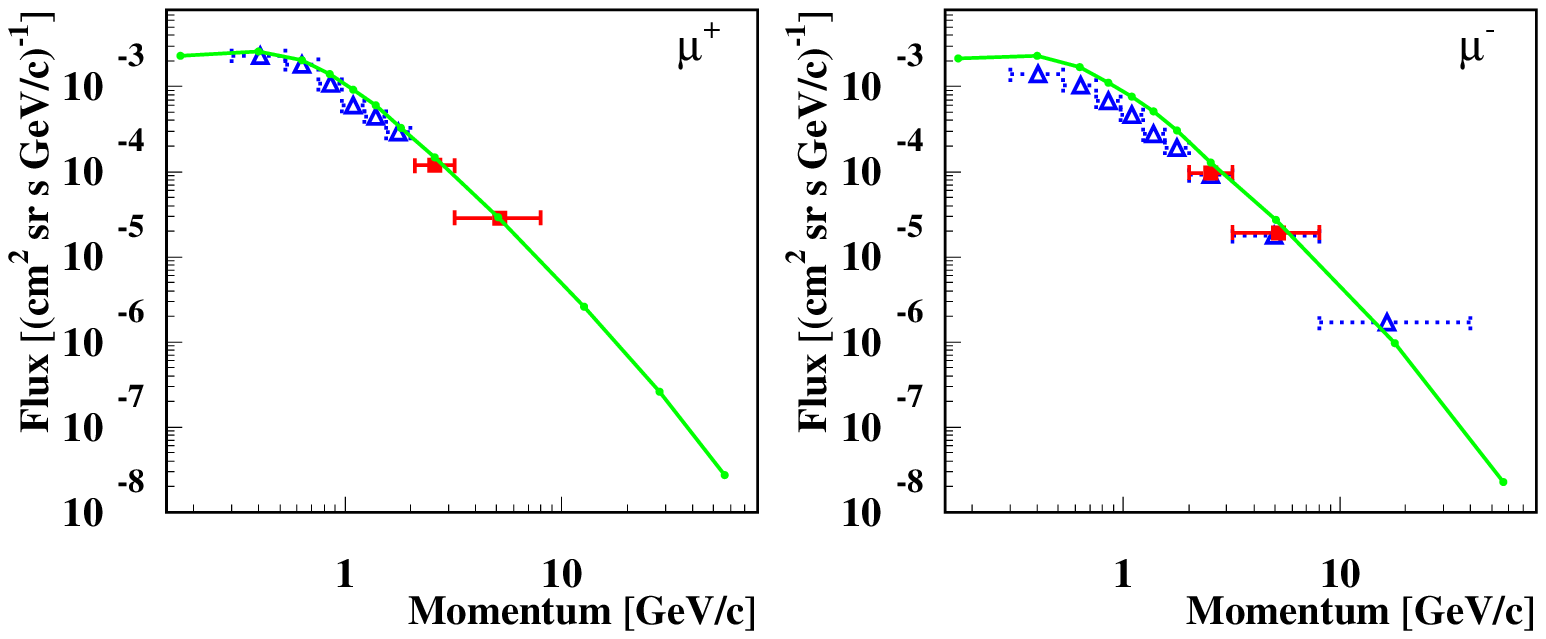}
\caption{ 
Absolute muon flux versus muon momentum at float altitude.
The solid squares 
(open triangles) correspond to CAPRICE98 (CAPRICE94)
data. The solid lines represent the AIRES simulations.
}
\label{fig:binfluxvsp9498float}
\end{center}
\end{figure}

\begin{figure}[p]
\begin{center}
\includegraphics{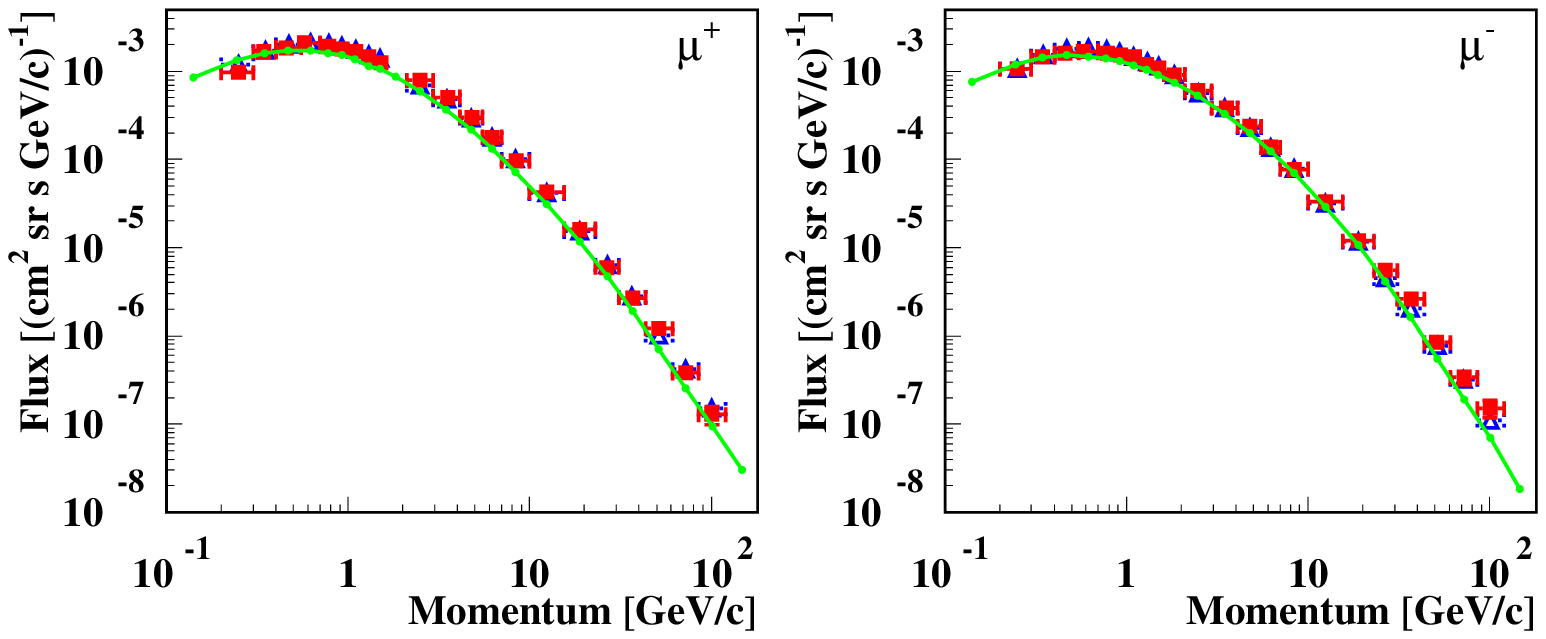}
\caption{ 
Absolute muon flux versus muon momentum at ground altitude.
The solid squares 
(open triangles) correspond to CAPRICE98 (CAPRICE97)
\cite{kremer}
data. The solid lines represent the AIRES simulations.}
\label{fig:binfluxvsp9498gnd}
\end{center}
\end{figure}

We have compared the fluxes obtained by Monte Carlo simulations
with the experimental data available at every altitude. 
Figures \ref{fig:binfluxvsp9498ascen},
\ref{fig:binfluxvsp9498float} and 
\ref{fig:binfluxvsp9498gnd} summarize the results obtained for the ascent, 
float and ground measurement phases, respectively. 
In these figures the fluxes of negative and positive muons are plotted
as functions of the muon momentum for every altitude considered.
In figure \ref{fig:binfluxvsp9498ascen}, that corresponds to the 
ascent phase of the CAPRICE98 flight, the fluxes corresponding 
to different altitudes have been multiplied by powers of 100, 
as indicated in the respective figure caption. 
The full lines correspond to the AIRES simulations; while the solid
squares represent CAPRICE98 data. When available, CAPRICE94 data
\cite{boe99b,muonpaper}
have been displayed as well (open triangles). Notice that the
  CAPRICE94 flight corresponds to a different geographical location,
  and a different time. Therefore the corresponding results,
  especially for low muon momenta, are not
  directly comparable to the simulations, that were performed taking
  into account the CAPRICE98 environment.

The shaded bands drawn together with each distribution illustrate how
the flux changes inside the atmospheric depth intervals of table
\ref{T:DepthIntervals}.  The varying width of these shaded areas can be
understood taking into account the behavior of the total muon flux
for varying atmospheric depth, represented in figure
\ref{fig:allabsfluxesvsxh}.  Near the top of the atmosphere, the muon
flux grows with $X$, and a visible difference between fluxes evaluated
at the minimum and maximum depth of each measurement interval exists
as made evident for all momenta in the highest levels plotted in figure
\ref{fig:binfluxvsp9498ascen}.  In these cases the lower (upper) curve
of the shaded region corresponds to the respective minimum (maximum).

On the other hand, in the cases of measurement intervals located close
to the ground level, the muon flux decreases with $X$, and therefore
the positions of the distributions corresponding to the minimum and
maximum depth of the respective interval interchange with respect to
the highest ones discussed in the previous paragraph. In these cases
there are no important variations at the high energy end of such
distributions.

For $X$ ranging from about 70 g/cm$^{2}$ to about 
200 g/cm$^{2}$ the muon flux remains approximately constant. In 
this case the curves corresponding to minimum, maximum and average
depth for a given level overlap.
As a consequence, any errors due to uncertainties in the altitude of
the balloon and/or the atmospheric model used in the simulations should
be very small at these intermediate depths.

\begin{figure}[p]
{\centering \begin{tabular}{cc}
\resizebox*{0.46\textwidth}{!}{\includegraphics{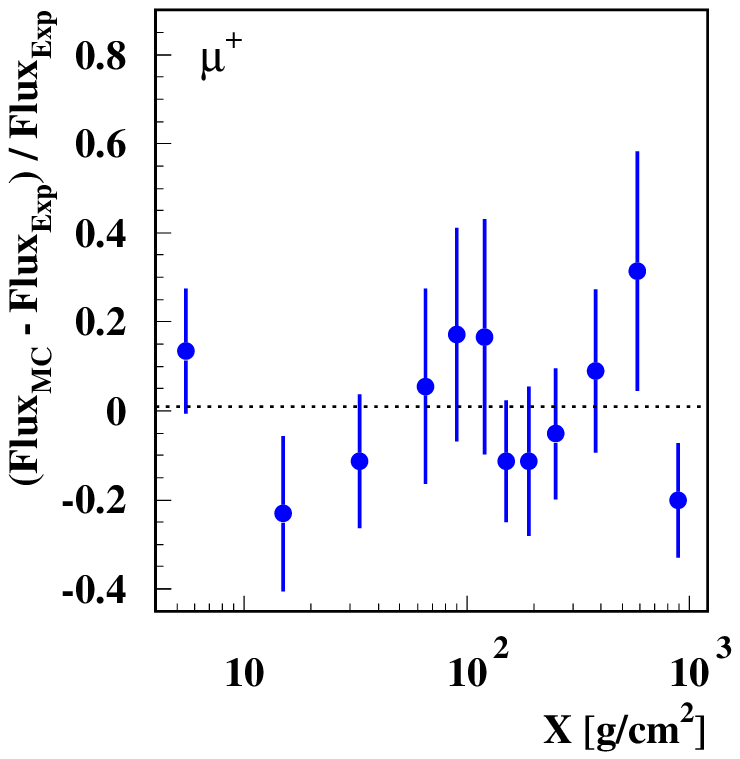}}
&\resizebox*{0.46\textwidth}{!}{\includegraphics{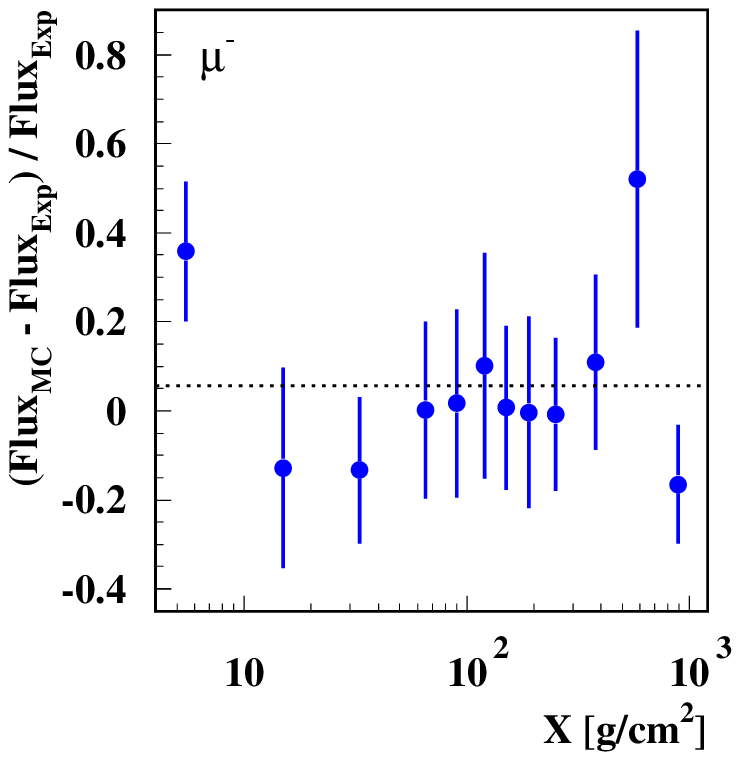}}
\\
\end{tabular}\par}
\caption{Relative difference simulation-experiment versus
atmospheric depth. Each point corresponds to a weighted 
average for all rigidity bins at the respective measurement
level. The dotted line represents the global average 
over all the considered altitudes.
}
\label{reldifc98vsx}
\end{figure}

The relative differences between simulated and experimental data,
are shown in figure \ref{reldifc98vsx} as function of the
atmospheric depth.
For each atmospheric depth, the relative difference is a weighted
average over momenta, where the weight is the inverse of the
relative error on each measured point. The error bars have been
calculated taking into account both the experimental and Monte Carlo
errors. To estimate these last ones we have taken into account only the
uncertainty derived from the uncertainty in the flux at TOA, whose
average over the energies with measurements is around
10\%. The Monte Carlo statistics is very large so statistical
fluctuations of mean values can be neglected. We have
not attempted in this work to do a detailed analysis of other
uncertainties that can affect the simulated data.

The relative differences are generally compatible with zero within
one standard deviation of the experimental flux.
 At float altitude the simulated flux for $\mu^{+}$ 
is 38 $\pm$ 18 \% above the measured one.
 At ground level the simulated flux is approximately
20 \% lower than the experimental one.

The global averages of the relative differences, represented in figure
\ref{reldifc98vsx} as dotted lines, are essentially zero (there is a
slight positive tendency of about 5 \% in the $\mu^-$ case, that is
compatible with zero well within $1\sigma$).

The ground level data is worth of a special analysis because in this case the
flux was measured with high statistics,
allowing for narrower momentum bins. The comparison between simulated
and experimental data indicate that the simulations predict a
flux that is smaller than the experimental one, especially for high
muon momenta, even if such difference is compatible with zero within
2$\sigma$. There are several possible reasons for these
differences, as already mentioned in the introduction. 
We discuss here some of our results.

The differences between simulated and experimental data
at ground, amounting to about 20\%, suggest
that there are systematic errors on that level 
since the statistical errors are very small.
The relative importance of primaries with energies
above 100 GeV (see figure \ref{fig:contribpegyfluxmuvsxh})
is larger for ground data than for data at higher altitudes,
suggesting that primary energy dependent systematic errors could have a non
negligible effect on the muon data at ground.
This will be the subject of a detailed study to appear later.

\begin{figure}[p]
\begin{center}
\includegraphics{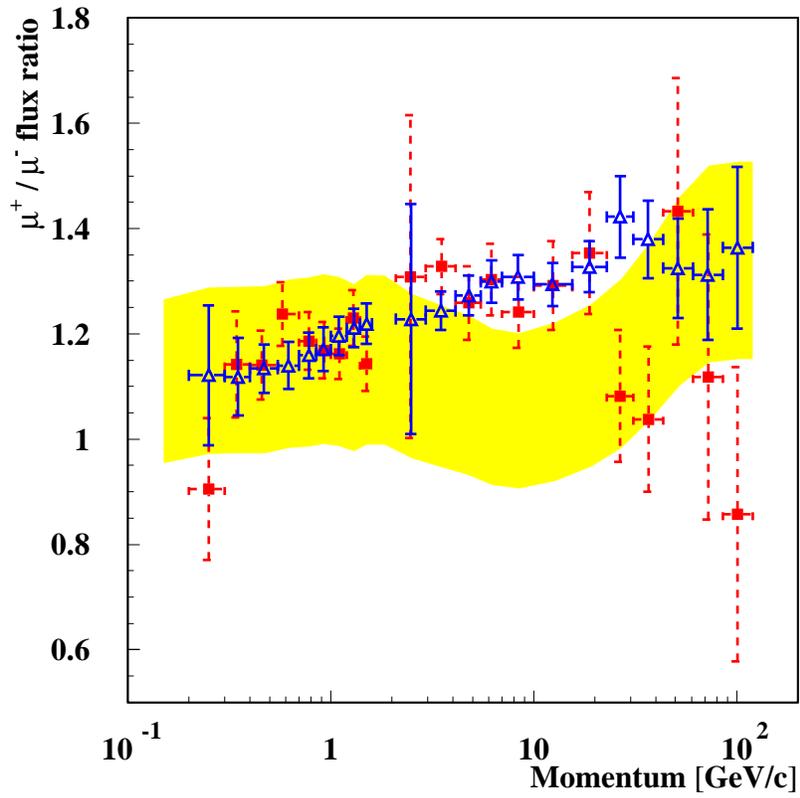}
\end{center}
\caption{$\mu^+/\mu^-$ flux ratio versus muon momentum, at ground
  altitude. The squares (triangles) correspond
  to CAPRICE98 (CAPRICE97)
  data. The shaded area corresponds to the simulations taking into
  account an estimation of the modelling errors.}
\label{mupmratiovsp}
\end{figure}

The study of the $\mu^+/\mu^-$ flux ratio can also provide important
information to test the models used in the simulations.  We consider
only the data at ground level where contamination corrections are
negligible, and the experimental data is abundant enough to give small
errors. In figure \ref{mupmratiovsp} the flux ratio at ground is
plotted versus muon momentum. The squares (triangles) correspond to
the CAPRICE98 (CAPRICE97) experimental data, while the shaded area
represents the simulation results including an approximate estimation
of modeling errors coming from different uncertainties (geomagnetic
effects, input flux, hadronic models, etc.). It is worthwhile
mentioning that the statistics of the CAPRICE97 data set is about 
four times larger than that of the CAPRICE98 case, resulting, as is
evident from the plot, in smaller error bars. When comparing the
results of the simulations with the experimental data, we find in
general an acceptable agreement, with relative differences always
located within 2$\sigma$.  However, in the approximate range 3-40
GeV/c the simulations give ratios that are about 10\% lower than the
experimental results.  Such differences are of the same order of
magnitude than the ones reported in a similar study performed using
various hadronic models \cite{wentzicrc}, giving a qualitative
indication of how much these models can change the simulated
$\mu^+/\mu^-$ flux ratio. A complete study of the impact of different
hadronic models, as well as other factors that could influence the
final simulated ratio is beyond the scope of this work, and
will be addressed in a future publication.

\begin{figure}[p]
\[
\begin{array}{cc}
\hbox{\includegraphics[width=0.4\textwidth]{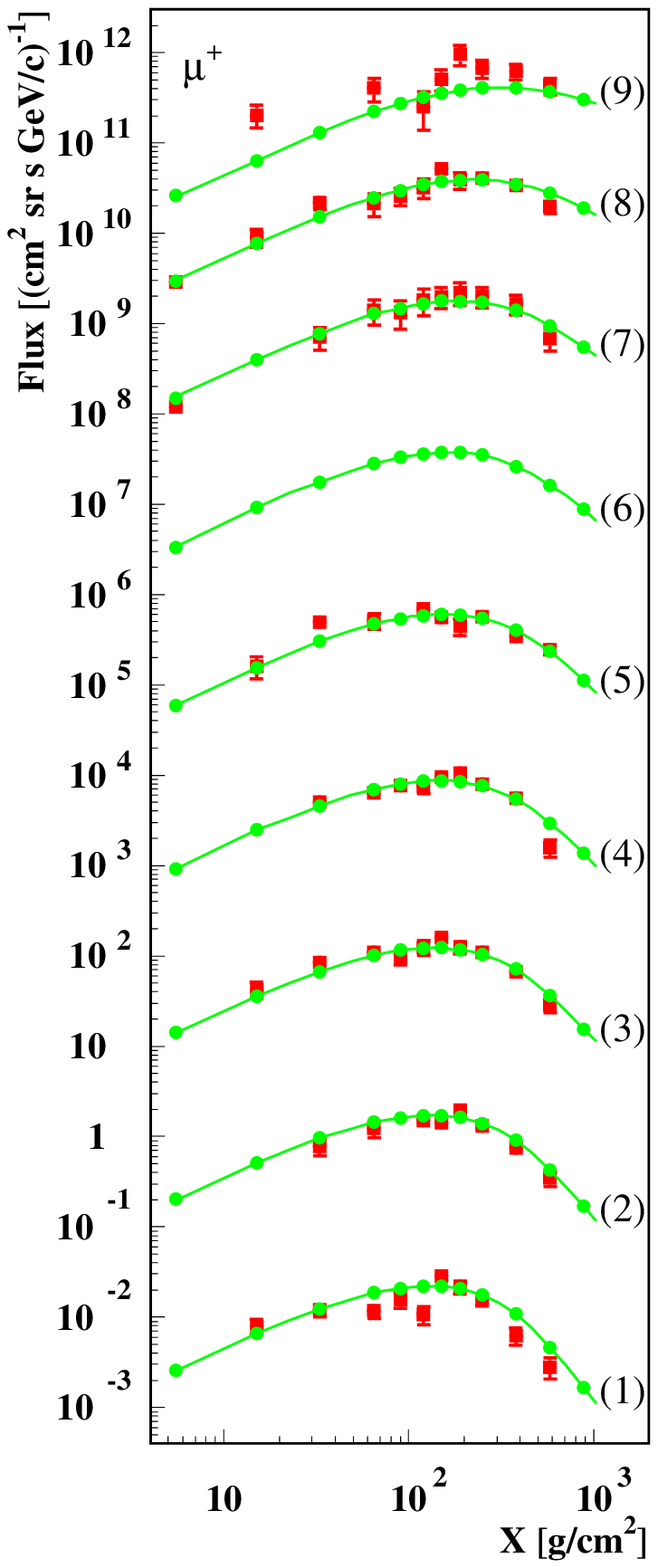}} &
\hbox{\includegraphics[width=0.4\textwidth]{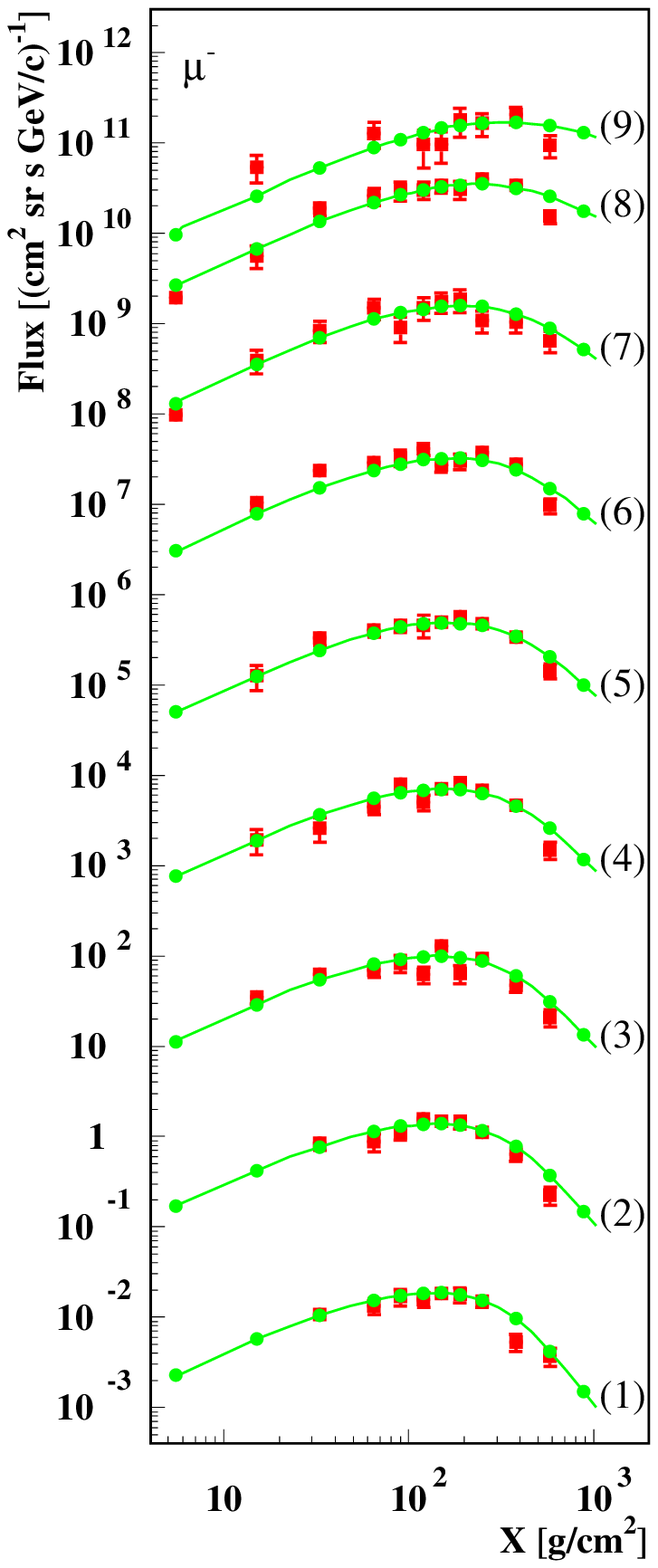}}
\end{array}
\]
\caption{
Absolute muon flux versus atmospheric depth. The squares represent
CAPRICE98 data, while the solid lines correspond to the simulations with
AIRES.
Different curves represent different momentum bins (in GeV/c); for
$\mu^+$ ($\mu^-$):
{\em (1)\/} 0.3 -- 0.53, (0.3 -- 0.53) scaled by 1.
{\em (2)\/} 0.53 -- 0.75, (0.53 -- 0.75)scaled by $10^2$.
{\em (3)\/} 0.75 -- 0.97, (0.75 -- 0.97) scaled by $10^4$.
{\em (4)\/} 0.97 -- 1.23, (0.97 -- 1.23) scaled by $10^6$.
{\em (5)\/} 1.23 -- 1.55, (1.23 -- 1.55) scaled by $10^8$.
{\em (6)\/} 1.55 -- 2.1, (1.55 -- 2.0) scaled by $10^{10}$.
{\em (7)\/} 2.1 -- 3.2, (2.0 -- 3.2) scaled by $10^{12}$.
{\em (8)\/} 3.2 -- 8, (3.2 -- 8) scaled by $10^{14}$.
{\em (9)\/} 8 -- 20, (8 -- 40) scaled by $10^{16}$.
}
\label{fig:depth}
\end{figure}

It is also interesting to study the change with altitude of the flux
of muons with a given momentum. In figure \ref{fig:depth} the absolute
fluxes of muons corresponding to each of the experimental momentum
bins are plotted as functions of the atmospheric depth.

Those curves indicate that there exists a very good agreement between
our simulations with AIRES and the experimental data in all the
considered cases, with no evident bias of any kind.
We also note that at each momentum bin both experimental and
simulated curves have their maxima approximately at the same altitude.

\section{Conclusions}
\label{S:Conclusions}

An in depth analysis of the absolute flux of muons at different
atmospheric depths has been performed. This includes a comparative
study of fluxes measured by the CAPRICE98 and CAPRICE94 experiment
with the corresponding simulations with AIRES. Additionally, the most
important correlations between the simulated observables and different input
parameters are discussed.

Many of the results presented in the previous sections indicate
clearly that the strategy followed to perform the simulations is
capable of producing acceptable results. We could also verify that the
following conditions cannot be simplified without altering to some
extent the simulated fluxes:
\begin{itemize}
\item Input fluxes that include H, He, and other heavy nuclei, with
  energies up to a maximum value of at least 10 TeV.
\item Full 3d propagation of particles.
\item Wide range of primary zenith angles, especially for simulation
  of full fluxes.
\end{itemize}

The fluxes simulated with AIRES present a very good agreement with the
corresponding ones measured experimentally, at all altitudes and
muon momenta considered. With the exception of a few isolated points,
the relative differences between simulated and experimental data are
always smaller than 20 \%. In most cases the relative differences are
compatible with zero within error bars. Global averages of relative
differences are technically zero, and there are no evident bias that
could indicate that the simulation algorithms are not performing
adequately.

Detailed comparisons between 
measured and simulated atmospheric muon data  are surely useful to tune
 the simulation
program but requires experimental measurements substantially more accurate
than the ones available at present. In this direction, the CAPRICE
re-flight project \cite{circella} constitutes a very important effort
capable of providing improved measurements that will allow the
production of more detailed comparative analysis of real and simulated
data.

\section{Acknowledgments}
One of the authors (S. J. S.) is indebted to Prof. Per Carlson and colleagues 
at the Department of Physics, KTH, Sweden, for their kind hospitality.
This work was partially supported by CONICET and Fundaci\'on Antorchas 
of Argentina.
P. H. thanks Roberto Liotta for useful
discussion and the Swedish Foundation for International
Cooperation in Research and Higher Education
for support.
E. M. was supported by the
Foundation BLANCEFLOR Boncompagni--Ludovisi, n\'ee Bildt.

%
%

\clearpage

\end{document}